\documentclass[useAMS,usenatbib]{mn2e}

\usepackage{psfig, epsf, epsfig}

\title[Origin of optically passive spirals]{Origin of optically passive spiral galaxies
with dusty star-forming regions: Outside-in truncation of star formation?}
\author[K. Bekki and   W. J. Couch]
{Kenji Bekki${}^1$\thanks{E-mail:
bekki@cyllene.uwa.edu.au}
and Warrick J. Couch${}^2$ \\
${}^1$ICRAR M468
The University of Western Australia
35 Stirling Hwy, Crawley
Western Australia, 6009 \\
${}^2$Centre for Astrophysics and Supercomputing, Swinburne University of
Technology, Hawthorn, Victoria 3122, Australia\\}

\begin{document}

\date{Accepted, Received 2005 February 20; in original form }

\pagerange{\pageref{firstpage}--\pageref{lastpage}} \pubyear{2005}

\maketitle

\label{firstpage}

\begin{abstract}

Recent observations have revealed that red, optically--passive spiral galaxies 
with little or no optical emission lines, harbour significant amounts of 
dust-obscured star formation. We propose that 
these observational results can be explained if the spatial distributions of 
the cold gas and star-forming regions in these spiral galaxies 
are significantly more compact than those in blue star-forming
spirals. Our numerical simulations show that if the sizes of 
star-forming regions in  spiral  galaxies with disk sizes
of $R_{\rm d}$ are $\sim 0.3R_{\rm d}$, such galaxies appear to
have lower star formation rates as well as higher degrees of dust extinction. 
This is mainly because star formation in these spirals occurs only in the 
inner regions where both the gas densities and metallicities are higher, 
and hence the dust extinction is also significantly higher.
We discuss whether star formation occurring preferentially  in the inner
regions of spirals is closely associated with the stripping of halo and disk gas
via some sort of environmental effect. We suggest that the 
``outside-in truncation of star formation'' is the key to a better understanding
of apparently optically--passive spirals with dusty star-forming regions.
\end{abstract}


\section{Introduction}

Since the discovery of significant numbers of galaxies in distant ($z\sim 0.2$--0.5)
clusters with a spiral morphology but with no apparent on-going star formation 
based on the absence of any emission lines in their optical spectra   
(e.g., Couch et al. 1994, 1998; Dressler et al. 1999; Poggianti et al. 1999), 
the origin of these so-called ``optically--passive'' spirals
(or ``k-type'' spirals) has received considerable attention both observationally 
and theoretically (e.g., Bekki et al. 2002; Goto et al. 2003; 
Yamauchi \& Goto 2004; Moran et al. 2006; 
Masters et al. 2010). For example,
Goto et al. (2003) found that such passive spirals are located 
anywhere between $1-10$ virial radii from the centres of clusters
and suggested that their formation is closely associated with 
cluster-related physical processes. In contrast, Masters et al. (2010) 
recently found that passive spirals 
exist preferentially in intermediate
density regimes, and that there are no obvious correlations between 
their physical properties and their environment.

\begin{table*}
\centering
\begin{minipage}{185mm}
\caption{The ranges of model parameters.}
\begin{tabular}{ccccccccc}
model no &
{$f_{\rm g}$ 
\footnote{The initial gas mass fraction of a  disk galaxy.}} &
{$R_{\rm g}$ ($\times R_{\rm d}$) 
\footnote{The initial gas disk size of a disk galaxy measured in units of
$R_{\rm d}$, where $R_{\rm d}$ (=17.5 kpc) is the initial (stellar) disk
size of the galaxy.}} &
{${\Sigma}_{\rm g,g}$ (${\rm M}_{\odot}$ pc$^{-2}$)
\footnote{The initial mean  surface gas density  of a disk galaxy
within $R_{\rm g}$.}}  &
{${\Sigma}_{\rm g,d}$ (${\rm M}_{\odot}$ pc$^{-2}$)
\footnote{The initial mean surface gas density  of a disk galaxy
within $R_{\rm d}$. }} &
{$\bar{a}_{\rm v}$ 
\footnote{The value of the extinction parameter $a_{\rm v}$
averaged over all time steps in a simulation.}}  &
{$\bar{a}_{\rm v,n}$ 
\footnote{The value of the extinction parameter $a_{\rm v}$
averaged over the last 0.1 Gyr in  a simulation.}}  &
{$\bar{b}_{\rm sf}$ (${\rm M}_{\odot}$ yr$^{-1}$) 
\footnote{The value of the star formation rate 
averaged over all time steps in a simulation.}}  &
{$\bar{b}_{\rm sf,n}$ (${\rm M}_{\odot}$ yr$^{-1}$) 
\footnote{The value of the star formation rate 
averaged over the last 0.1 Gyr  in a simulation.}}  \\
Model 1 &  0.05 & 0.3 & 37.42 & 3.37 & 0.306 & 0.186 & 8.55 & 3.37 \\
Model 2 &  0.20 & 1.0 & 12.47 & 12.47 & 0.067 & 0.056 & 15.17 & 10.90 \\
\end{tabular}
\end{minipage}
\end{table*}

A further important and yet puzzling observational result is that some of the 
passive spirals contain significant amounts of {\it obscured} star formation 
(e.g., Wolf e al. 2005; Wilman et al. 2008; Wolf et al. 2009).
The star formation rates in these cases are a factor of $\sim 4$
lower than those in blue spirals with the same mass. More specifically,   
the ratio of the star formation rate inferred from their infrared emission 
(SFR$_{\rm IR}$) to that inferred from their UV emission (SFR$_{\rm UV}$)
is typically a factor of $\sim 3$ larger than that for blue spirals
(Wolf et al. 2009), implying that the passive spirals have a
significantly higher ($\times\sim 2$) level of dust extinction. 
However, it remains unclear how and when such dust extinction 
occurs in spiral galaxies when their star formation rates are
significantly lower.

The purpose of this paper is to show how
optically--passive spirals with lower yet substantial
star formation rates can have higher degrees of dust extinction,
based on numerical simulations of star-forming disk galaxies.
In particular, we demonstrate that if the sizes of actively 
star-forming regions ($R_{\rm sf}$) in galaxies with disk sizes 
of $R_{\rm d}$  are significantly more compact than $R_{\rm d}$,
then they will exhibit both lower star formation rates and
heavier dust extinction.
We assume that $R_{\rm sf}$ varies amongst spiral galaxies in the 
local and distant universe, and thus we treat $R_{\rm sf}$
as a free parameter in this study. This assumption is consistent 
with the recent observations of Koopmann \& Kenny (2004) which showed
that isolated spirals as well as those in the Virgo cluster 
are extremely diverse in the radial distribution and extent of 
their H$\alpha$ emission (from star-forming regions). 
We discuss how $R_{\rm sf}$ can be more compact in 
some star-forming passive spirals in \S 4.

\section{The model}

We used the latest version of GRAPE
(GRavity PipE, GRAPE-7) -- which is the special-purpose
computer for gravitational dynamics (Sugimoto et al. 1990) -- 
in order to investigate the chemodynamical evolution of star-forming disk galaxies. 
We have revised our original GRAPE-SPH code (Bekki 2009)
for galaxy-scale hydrodynamical evolution 
so that we can investigate chemical 
evolution and star formation processes of disk galaxies; the details
of this new code will be described in future papers
(e.g., Bekki et al. 2010).

The masses of the disk, bulge ,and dark halo components of our
model galaxy are represented by $M_{\rm d}$, $M_{\rm b}$, and $M_{\rm dm}$, respectively.
The mass ratio of $M_{\rm dm}$ to $M_{\rm d}$
was fixed at 16.7 for all  of the present models 
so that the models can mimic the mass distribution
of the Galaxy with the total mass of $\sim 10^{12} {\rm M}_{\odot}$
(e.g., Evans \& Wilkinson 2000).
We adopted an NFW halo density distribution (Navarro, Frenk \& White 1996) 
suggested from CDM simulations:
\begin{equation}
{\rho}(r)=\frac{\rho_{0}}{(r/r_{\rm s})(1+r/r_{\rm s})^2},
\end{equation}
where  $r$, $\rho_{0}$, and $r_{\rm s}$ are
the spherical radius,  the characteristic  density of a dark halo,  and the
scale
length of the halo, respectively.
We adopted $c=7.8$ (the ratio of $r_{\rm vir}$ to $r_{\rm s}$,
where $r_{\rm vir}$ is the virial radius)  that is reasonable 
for the total masses ($\sim 10^{12} {\rm M}_{\odot}$)
investigated in the present study.

The radial ($R$) and vertical ($Z$) density profiles of the disk
(with size $R_{\rm d}$) were
assumed to be proportional to $\exp (-R/R_{d,0})$, with scale
length $R_{d,0}$ = 0.2$R_{\rm d}$, and ${\rm sech}^2 (Z/Z_{d,0})$, with scale
length $Z_{d,0}$ = 0.04$_{\rm d}$ in our units, respectively; both the stellar 
and gaseous disks follow this exponential distribution.
In addition to the rotational velocity caused both by the gravitational fields
of 
the disk and dark halo components,
the initial radial and azimuthal velocity dispersions was assigned to the disk 
component according to the epicyclic theory with Toomre's parameter $Q = 1.5$.

The mass ratio of the bulge to the disk  ($f_{\rm b}$) and
the scale length ($R_{\rm b,0}$) of the stellar bulge represented by the Hernquist
profile were fixed at 0.167 and 0.04$R_{\rm d}$,
respectively, for all models, which is consistent with that of the bulge
model of the Galaxy.
In the present study, we describe only the results of the models
with $M_{\rm d}=6 \times 10^{10} {\rm M}_{\odot}$
(thus $M_{\rm b}=1.0 \times 10^{10} {\rm M}_{\odot}$)
and $R_{\rm d}=17.5$ kpc (thus $R_{\rm b,0}=0.7$ kpc).

The cold interstellar medium (ISM) was distributed within $R_{\rm g}$ 
and modeled using SPH particles. The mass fraction of the isothermal
ISM in the disk ($f_{\rm g}$) was assumed to be a free parameter. 
The initial temperature  of the ISM was assumed to be $10^4$ K.
We adopted the same method as that used in  
Bekki \& Chiba (2005) for determining the radial dependence of gas mass
fraction in a disk for a given $f_{\rm g}$. 
We adopted the same chemical evolution model as those used in Bekki \& Chiba (2005)
and the chemical yield and the return parameter were set to 0.02 and 0.3,
respectively. The star formation was assumed to follow the
Schmidt law (Schmidt 1959) with an exponent of 1.5.
Kinetic energy of $10^{51}$ erg per supernova is given to 
the ISM immediately after star formation occurs.

Friel (1995) has derived the metallicity ($Z$) gradient of the Galactic stellar disk 
based on the ages and metallicities that are estimated for the Galactic open
clusters. We therefore allocated metallicity to each disk star
according to its initial position as follows:
\begin{equation}
{\rm [m/H]}_{R} = {\rm [m/H]}_{\rm d, 0} 
+ {\alpha}_{\rm d} \times R,
\end{equation}
where $ {\rm [m/H]}_{\rm d, 0}$ is the central metallicity.
If we adopt plausible values of $-0.091$ 
for the slope ${\alpha}_{\rm d}$ (Friel 1995)
and the central value of $0.48$ for ${\rm [m/H]}_{\rm d, 0}$,
the mean metallicity of the disk is $0.0$ in [Fe/H].

In order to more quantitatively estimate dust extinction around each 
individual star in star-forming disk galaxies, 
we introduced a dimensionless parameter, $a_{\rm v}$,
which measures the degree of dust extinction for each new 
$i$-th stellar particle as follows: The dust extinction
at wavelength $\lambda$  ($A_{\lambda}$) around a star is described as
follows (e.g., Spitzer 1978):
\begin{equation}
A_{\lambda}=-2.5\log \frac{F_{\nu}}{F_{\nu}(0)}=1.086N_{\rm d}Q_{\rm e}
{\sigma}_{\rm d},
\end{equation}
where $F_{\nu}$, $F_{\nu}(0)$, $N_{\rm d}$, $Q_{\rm e}$, and ${\sigma}_{\rm d}$
are the observed radiative flux, the radiative flux in the absence
of extinction, the  column density (per cm$^{2}$) along the line of sight,
the  dimensionless extinction efficiency factor,
and the geometrical cross section of a dust particle.

We assumed here that ${\sigma}_{\rm d}$ and the dimensionless factor
$Q_{\rm e}$ are constant for all the models considered in this present study.
Furthermore, previous models have shown that the dust mass of the Galaxy
is linearly proportional to the total mass of the heavy
elements (e.g., Dwek 1998).
Therefore we took the degree of dust extinction ($A_{\rm v}$)
around a new star to be proportional to $N_{\rm d}$,
which in turn is proportional to ${\rho}_{\rm g} Z$,
where ${\rho}_{\rm g}$ and $Z$ are the 3D gas density and the gaseous
metallicity around the star, respectively.
Thus we defined the dust extinction parameter 
($a_{\rm v, \it i}$) for each individual $i$-th new
star as follows:
\begin{equation}
a_{\rm v, \rm i}={\rho}_{\rm g, \it i} Z_{i},
\end{equation}
where ${\rho}_{\rm g, \it, i}$ and $Z_{i}$ are the
3D gas density and metallicity around the star.
We here consider that
the present model enables us to discuss
the importance of initial gaseous distributions in determining
the mean dust extinction of a galaxy without using
our previous fully consistent model considering 3D dust distributions
(Bekki \& Shioya 2000).

We mainly investigated the time evolution
of the mean $a_{\rm v}$ of new stars 
and the star formation rate ($b_{\rm sf}$) over a 0.28\,Gyr  
time interval in each simulation.
We estimated the mean values of $a_{\rm v}$ and $b_{\rm sf}$ in star-forming
disk galaxies  over the last 0.1 Gyr (${\bar{a}}_{\rm v,n}$ and 
${\bar{b}}_{\rm sf,n}$, respectively)
and adopted them as reasonable indicators of the amount of dust
extinction and global star formation for the galaxies.
We consider that these mean values are better than the mean 
$a_{\rm v}$ and $b_{\rm sf}$  estimated for all time steps
(${\bar{a}}_{\rm v}$ and ${\bar{b}}_{\rm sf}$,
respectively), because strong starbursts can occur {\it initially}
in the inner regions of the disks for most models, 
owing to very high gas densities there. Thus our estimates pertain to
those times when star formation and chemical evolution proceeds
steadily in the disks.

The two key parameters in the present study are $f_{\rm g}$ and $R_{\rm g}$
which control the sizes of the star-forming regions ($R_{\rm sf}$).
Although we have run numerous models with different $f_{\rm g}$ and 
$R_{\rm g}$ values, we show mainly the results for two representative 
models, the key parameter values for which are given in Table 1. 
This is mainly because these two comparative models show most clearly
how the initial size of the gas disk is important
in determining the mean star formation rate and 
the mean degree of dust extinction in a spiral galaxy. 
These two models are hereafter referred to as Model 1, which refers
to a passive spiral with a lower star formation rate and higher level
of dust extinction, and Model 2, which refers to a blue spiral with
a higher star formation rate and lower level of dust extinction.
We also briefly describe the dependences of mean $a_{\rm v}$ on model
parameters ($f_{\rm g}$ and $R_{\rm g}$). The mass and scale resolutions
of the present simulations are $3 \times 10^5 {\rm M}_{\odot}$
and 193\,pc, respectively,
so that we can estimate $a_{\rm v}$ for local gaseous regions
($\sim 100$ pc).

Finally, we note that we are unable to discuss whether the simulated 
disks exhibit k-type spectra, because the present new chemodynamical
simulation does not output spectrophotometric information. 
We will address this important point in our future papers using
chemodynamical simulations with spectroscopic synthesis
code like those in our previous work (Bekki et al. 2001).


\begin{figure}
\psfig{file=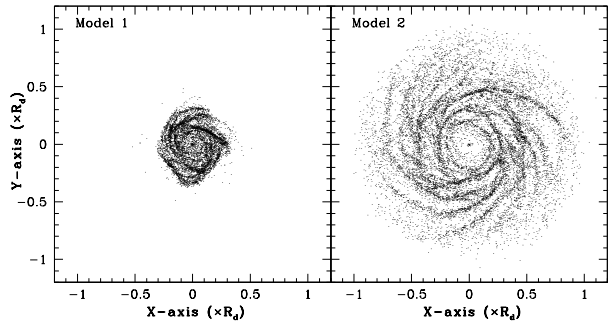,width=8.0cm}
\caption{
The distribution of new stars with ages less than 0.28 Gyr projected
onto the $x$-$y$ plane (i.e., the disk plane) for the two representative
models, Model 1 (left) and Model 2 (right). 
}
\label{Figure. 1}
\end{figure}

\begin{figure}
\psfig{file=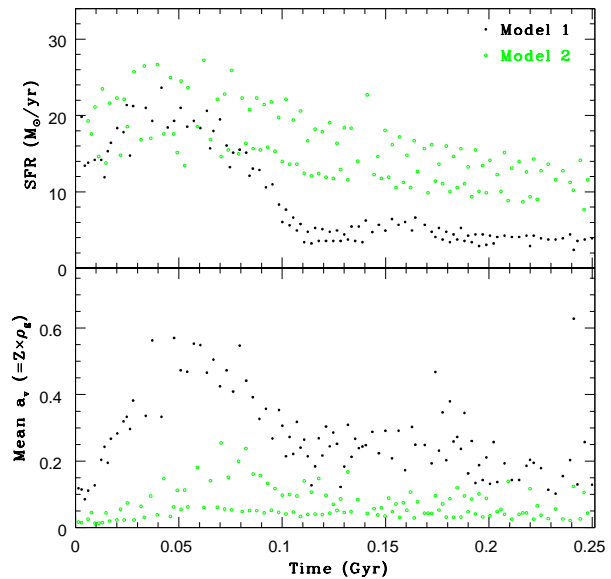,width=8.5cm}
\caption{
Time evolution of star formation rates (SFRs, upper) and mean $a_{\rm v}$
(extinction parameter, lower) for 0.28 Gyr in  Model 1 (filled black) and 2 
(open green).
}
\label{Figure. 2}
\end{figure}

\begin{figure}
\psfig{file=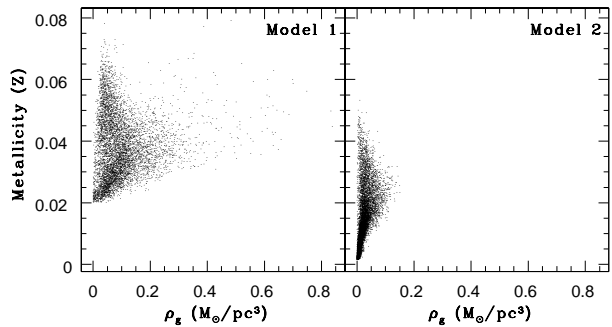,width=8.5cm}
\caption{
Plots of new stellar particles onto the  $Z-{\rho}_{\rm g}$ plane
for Model 1 (left) and 2 (right). Here the metallicity ($Z$) and
the local 3D gas density (${\rho}_{\rm g}$)
around each individual new stellar particle is plotted.
}
\label{Figure. 3}
\end{figure}

\begin{figure}
\psfig{file=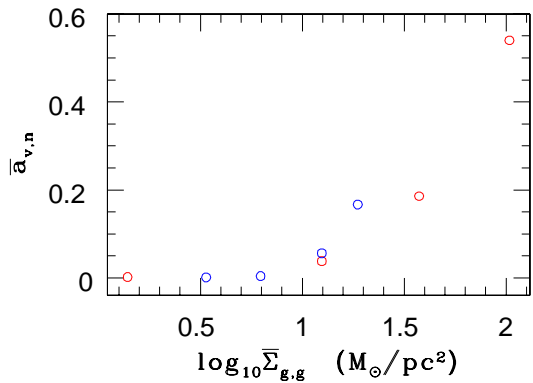,width=8.0cm}
\caption{
The dependences of  $\bar{a}_{\rm v,n}$ 
(mean dust extinction for new stars with ages less than 0.1 Gyr)
on ${\Sigma}_{\rm g,g}$
(initial mean gas density within $R_{\rm g}$) for 8 different models.
The models with $R_{\rm g}=0.3R_{\rm d}$ and $R_{\rm g}=R_{\rm d}$
are shown in red and blue, respectively. 
}
\label{Figure. 4}
\end{figure}

\section{Results}

Figure 1 shows that owing to the initially different
gaseous distributions, the distributions of newly-formed stars 
are significantly different between the two models, Model 1 and 2.
In Model 1, star formation can only proceed in the inner regions
where both gas densities and metallicities are high. 
Owing to a larger gas mass fraction, new stars can be formed across 
the entire region of the disk in Model 2; star formation can occur 
not only in the inner regions with higher gas densities and metallicities, 
but also in the outer regions, even though the gas densities and 
metallicities are lower. The mean star formation rate in Model 2 
($10.9 {\rm M}_{\odot} {\rm yr}^{-1}$ for the last 0.1 Gyr)
becomes  significantly
higher than that in Model 1 ($3.4  {\rm M}_{\odot} {\rm yr}^{-1}$)
owing to the initially larger $f_{\rm g}$.
The spiral-like structure delineated by the 
very young stars with ages less than 0.28 Gyr
in Model 2, suggests that 
well-defined large spiral structures
can be more clearly seen in disks with globally active star-forming
regions.

In Figure 2 the time evolution of the mean star formation rate and
extinction for Models 1 and 2 are shown. Here it can be clearly seen 
that the total star formation rate finally becomes as low as 
$\sim 3 {\rm M}_{\odot} {\rm yr}^{-1}$ in Model 1, 
whereas its mean $a_{\rm v}$ becomes a factor of $\sim 3$ higher
than that of Model 2 at $T=0.28$ Gyr.
The origin of this higher mean $a_{\rm v}$
in the disk of Model 1 can be explained via reference to Figure 3. 
This plots the 3D gas densities (${\rho}_{\rm g}$) 
and metallicities ($Z$) around the new stars formed within
0.28\,Gyr in Models 1 and 2. Here it can be clearly seen that
these two quantities are, on average, systematically higher in
Model 1 than in Model 2. This is due to the fact that in
Model 1 star formation occurs preferentially in the inner regions
with higher gas densities and metallicities, whereas in
Model 2 it occurs even in the lower-density and lower-metallicity 
outer regions of the disk. These results suggest that the observed  
higher dust extinction in passive spirals with lower star formation rates
might be closely associated with more centrally-concentrated gas 
distributions within them.

In the present study, ${\Sigma}_{\rm g,g}$
depending both on $f_{\rm g}$ and $R_{\rm g}$
is a key parameter that can determine ${\bar{a}}_{\rm v, n}$.
Figure 4 shows that the mean $a_{\rm v}$ for the very young stars formed 
within the last 0.1 Gyr (${\bar{a}}_{\rm v, n}$) depends on the initial 
mean gas densities within $R_{\rm g}$ (${\Sigma}_{\rm g,g}$) in such a 
way that ${\bar{a}}_{\rm v, n}$ is higher in models with higher 
${\Sigma}_{\rm g,g}$. This figure also shows that models with 
$R_{\rm g}=R_{\rm d}$ can
show higher ${\bar{a}}_{\rm v, n}$,
if $f_{\rm g} \ge 0.3$ 
(corresponding to ${\Sigma}_{\rm g,g} \ge 18.71 {\rm M}_{\odot} 
{\rm pc}^2$). 
Furthermore, Figure 4 shows that ${\bar{a}}_{\rm v, n}$
can be quite low in models with $R_{\rm g}=0.3R_{\rm d}$
if $f_{\rm g}$ is lower ($\le 0.02$): 
the star formation rates
are  also quite low in these models ($< 1 {\rm M}_{\odot} {\rm yr}^{-1}$).
This result suggests that passive spiral galaxies need to have a certain 
minimum amount of gas centrally concentrated in their disks if
they are to show both lower yet substantial star formation rates 
and higher dust extinction relative to normal blue spirals.

We have shown that if star-forming regions are very strongly
concentrated in the inner regions of spiral galaxy disks  
($R_{\rm g} \le 0.2R_{\rm d}$), then they appear to have
rather low star formation rates 
($<1 {\rm M}_{\odot} {\rm yr}^{-1}$) 
and high levels of dust extinction (${\bar{a}}_{\rm v, n} > 0.15$).
For example, the model with $f_{\rm g}=0.007$ and $R_{\rm g}=0.1R_{\rm d}$
shows ${\bar{b}}_{\rm sf,n}=0.85 {\rm M}_{\odot} {\rm yr}^{-1}$ 
and ${\bar{a}}_{\rm v, n}=0.30$.
This result implies that if most of the gas in gas-poor disk galaxies
(with $f_{\rm g} <0.01$)
can be fueled to the nuclear regions and consumed rapidly there
owing to some physical mechanism (e.g., galaxy-galaxy interaction),
then such disk galaxies can be identified  as passive spirals
with nuclear star formation
with higher  degrees of dust extinction.

\section{Discussion and conclusions}

If the scenario presented here (preferred star formation in
inner regions of galaxies) is correct, this begs the question
as to how the gas within disk galaxies might be truncated in
this way in the course of their evolution.  
Previous numerical simulations show that ram pressure stripping
by the hot intracluster medium can remove, quite efficiently, 
gas from the outer parts of spiral disks, so that their gas disk
becomes much more compact than their stellar disk  
(e.g., Abadi et al. 1999; Kronberger et al. 2008).
Furthermore, recent numerical simulations have shown that
ram pressure stripping of halo gas, which is an important
source of fuel for star formation in galactic disks, is more 
efficient in the outer parts of halos of disk galaxies 
in groups and clusters of galaxies (e.g., Bekki 2009).
Thus it is possible that truncated gas disks can be
formed as a result of halo and disk gas stripping, particularly 
in group and cluster environments.

Recently, the Galaxy Zoo project has revealed a large optical
bar fraction in red spirals at low redshift ($70\pm 5 \%$ versus $27\pm 5 \%$ 
for blue spirals),
and thus suggested that stellar bars are responsible for
the truncation of star formation in this subset of the spiral 
galaxy population (Masters et al. 2010). Tidal interactions between
galaxies can trigger the formation of bars and consequently transfer rapidly
disk gas into the inner regions of the galaxies (e.g., Noguchi 1988).
Therefore, the bars in disk galaxies can 
change the spatial distribution of gas such that the distribution
can be much more centrally concentrated.
Thus it could well also be possible that the origin of the proposed truncated
gas disks has something to do with dynamical action of stellar bars in 
disk galaxies.

An important and possibly testable prediction of the scenario presented here
is that any emission associated with the star formation in optically--passive
red spirals should be much more compact that that associated with star 
formation in blue spiral galaxies. On the other hand, the star formation rate per
unit area (i.e., star formation density measured in units of  
${\rm M}_{\odot}$ yr$^{-1}$ kpc$^{-2}$) for these two different types of
spirals should not be so different, because the star formation densities
within the inner regions of passive spirals are expected to be as high
as those in blue spirals. What is needed to test this is an emission line
that traces star formation (and its rate) and is not heavily affected
by dust extinction. Here, the H$\alpha$ line probably holds the most promise
in terms of being the optical line least affected by dust extinction and
one that can be readily mapped spatially and out to high redshifts via 
high resolution imaging and integral field unit spectroscopy.  

As commented on above, a shortcoming of this study is its inability to
show that disk galaxies with centrally-concentrated star formation
exhibit passive k-type spectra.  Although previous theoretical studies based on
one-zone models and numerical simulations showed that e(a) and a+k/k+a spectra can be formed
in dusty star-forming galaxies (e.g., Shioya \& Bekki 2000; Bekki et al. 2001;
Shioya et al. 2002), they did not clearly show k-type spectra can be formed from
dusty star-forming galaxies. Thus it is crucially important that as a next step we conduct 
further numerical simulations that include spectrophotometric modeling which
allow us to predict the spectroscopic signature associated with 
star formation that proceeds in the inner regions of disk galaxies.

If the star-forming regions of red passive spirals are as spatially 
extended as those in blue star-forming spirals, then it will be
necessary to consider alternative scenarios to the one presented here.
One such possibility worthy of brief mention here is that the upper-mass 
cutoff ($m_{\rm upp}$) of the initial mass function (IMF) is significantly 
smaller (e.g., $<20 {\rm M}_{\odot}$) in passive spirals.  In this truncated 
IMF scenario, there are few or no massive O stars that can ionize 
the ISM (i.e., $>20 {\rm M}_{\odot}$), such that (i)\,optical emission lines  
are very weak, and (ii)\,dust in the ISM can obscure star formation quite 
efficiently due to there being little destruction of dust by ionizing photons.
This truncated IMF scenario has observational support through being able
to explain the UV and H$\alpha$ properties of low surface brightness galaxies 
with low star formation rates (Meurer et al. 2009).
More quantitative investigation based on numerical simulations of disk galaxy 
evolution with non-universal IMFs are required to test
the viability of this scenario.

Recent observational studies of distant galaxies based on 
{\it Spitzer} 24$\mu$m photometry and
optical imaging by the {\it Hubble Space Telescope} 
have revealed the dusty nature of red galaxies and have provided
new clues to the possible gradual truncation  of galactic star formation 
in different environments
(e.g., Gallazzi et al. 2009; Wolf et al. 2009).
The present study suggests that truncation of star formation
can occur more dramatically in the outer parts of disk galaxies,
where environmental processes  (e.g., tidal and ram pressure stripping)
can be more effective. It also suggests that
the possible inner dusty star-forming regions in passive spirals
would be due to ``outside-in truncation of star formation'' in the course
of disk galaxy evolution, in particular, in groups and clusters of galaxies.

\section{Acknowledgment}
We are grateful to the  referee Christian Wolf  for valuable comments,
which contribute to improve the present paper.
KB  and WJC all acknowledge the financial support of the
Australian Research
Council throughout the course of this work. Numerical computations
reported here were carried out both on the GRAPE system at the
University of Western Australia  and on those kindly made available
by the Center for computational astrophysics
(CfCA) of the National Astronomical Observatory of Japan.

\end{document}